\def\e{{\rm e}}
\def\Z#1{_{\lower2pt\hbox{$\scriptstyle#1$}}}

\documentclass{ws-mpla}

\begin{document}

\markboth{Ishwaree P. Neupane} {GAUSS-BONNET COSMOLOGIES}

%
\catchline{}{}{}{}{}
%

\title{CONSTRAINTS ON GAUSS-BONNET COSMOLOGIES}

\author{\footnotesize Ishwaree P. Neupane}

\address{Department of Physics and Astronomy, University of Canterbury,\\
Private Bag 4800, Christchurch 8020, New Zealand\\
$^*$E-mail: ishwaree.neupane@canterbury.ac.nz}

\maketitle

\pub{Received (10 Nov 2007)}{Published (Day Month Year)}

\begin{abstract}

The modified Gauss-Bonnet gravity can be motivated by a number of
physical reasons, including: the uniqueness of a gravitational
Lagrangian in four and higher dimensions and the leading order
$\alpha^\prime$ corrections in superstring theory. Such an
effective theory of scalar-tensor gravity has been modeled in the
recent past to explain both the initial cosmological singularity
problem and the observationally supported cosmological
perturbations. Here I present an overview of the recent
developments in the use of modified Gauss-Bonnet gravity to
explain current observations, touching on key cosmological and
astrophysical constraints applicable to theories of scalar-tensor
gravity. The Gauss-Bonnet type modifications of Einstein's theory
admits nonsingular solutions for a wide range of scalar-curvature
couplings. It also provides plausible explanation to some
outstanding cosmological conundrums, including: the transition
from matter dominance to dark energy and the late time cosmic
acceleration. The focus is placed here to constrain such an
effective theory of gravity against the recent cosmological and
astrophysical observations.

\end{abstract}


\section{Introduction}

Einstein's general relativity has been very successful as a
classical theory of gravitational interactions, especially, in a
non-accelerating (or non-expanding) spacetime. In a cosmological
background, the theory predicts spacetime singularities, so its
modification is inevitable at high energy scales. Further the
recently observed accelerating expansion of the
universe~\cite{Supernova} provides some insight to the possibility
that general relativity together with ordinary matter and
radiation, described by the standard model of particle physics,
cannot fully explain the current observations. The question arises
because the current observations~\cite{WMAP03} require in the
fabric of the cosmos the existence of a dark energy component of
magnitude about $73\%$, which does not `clump' gravitationally.
Another $23\%$ of the mass-energy is in the form of mysterious
non-baryonic dark matter, which `clumps' gravitationally.

The bulk of the universe appears to be dark energy and dark
matter. So far there is no fully consistent explanation of these
energy components supported by a fundamental theory. The main
focus of this meeting is obviously to update our knowledge on DARK
matter and DARK energy searches and the physics behind these. The
focus of my presentation will be on a possible resolution of dark
energy problem within some string-inspired theories of
scalar-tensor gravity.

\section{Accelerating universes and string-inspired models}

The discovery that the expansion of the universe is currently
accelerating is among the most tantalizing (and perhaps most
mysterious) of recent times. Evidence in favour of this
accelerated expansion (caused by putative dark energy) has led to
a continued interest in scenarios that propose modifications to
Einstein's general relativity. The proposals are of differing
origins as well as motivations, some are based on theories of
higher-dimensional gravity and others on consideration of one or
more fundamental scalar fields and their interactions with
higher-order curvature terms. Both these ideas are well motivated
by supergravity and superstring theories, which incorporate
Einstein's theory in a more general framework. There are several
theoretical motivations to incorporate string theory into
cosmological model building. Notably, gravitational interactions
mediated by scalar fields, together with the standard graviton,
are the best motivated alternatives~\cite{Veneziano:2000} to
general relativity, as they provide a mathematically consistent
framework to test the various observable predictions of higher
dimensional theories of gravity, such as, brane inflation.

Typically the low energy limit of string theory or supergravity
features scalar fields and their couplings to a unique combination
of the three quadratic scalars $R^2$, $R_{\mu\nu}R^{\mu\nu}$ and
$R_{\mu\nu\rho\lambda}R^{\mu\nu\rho\lambda}$, composed of the
scalar curvature, the Ricci and Riemann tensors:
$$ {\cal R}^2\equiv R^2-4 R_{\mu\nu} R^{\mu\nu} +
R_{\mu\nu\rho\lambda} R^{\mu\nu\rho\lambda}, $$ known as the
Gauss-Bonnet term. This term arises, almost universally, in all
versions of string theory as the leading order $\alpha^\prime$
correction. An illustrative example is the following
four-dimensional heterotic superstring model which describes the
dynamics of graviton, dilaton $S$ and the common (volume) modulus
field $T$, arising from a compactification of 10D heterotic
superstring theory on a symmetric 6D
orbifold~\cite{Antoniadis:1992}:
\begin{equation}
{\cal L}_{\rm grav}= {\cal L}_{0}+ {\cal L}_{1},
\end{equation}
where the string tree-level Lagrangian ${\cal L}_0$ is
\begin{eqnarray}
{\cal L}_{0} =  \frac{R}{2\kappa^2}-\zeta \frac{2\Delta S \Delta
\bar{S}}{(S+\bar{S})^2}- \gamma \frac{2\Delta T \Delta
\bar{T}}{(T+\bar{T})^2} +\frac{1}{8}({\rm Re} S) {\cal R}^2
+\frac{1}{8} ({\rm Im} S)\,R\tilde{R},
\end{eqnarray}
while the modulus $T$ dependent Lagrangian at the one loop level
is
\begin{equation}
{\cal L}_{1} = \Delta (T,\bar{T})\,{\cal R}^2 -i\Delta
(T,\bar{T})\,R \tilde{R},
\end{equation}
where $\kappa$ is the inverse Planck mass $M_{P}^{-1}=(8\pi
G_N)^{1/2}$, $G_N$ is Newton's constant, $\zeta$ and $\gamma$ are
numerical constants, $R\tilde{R} \equiv g^{-1/2}
\epsilon^{\mu\nu\rho\lambda} R_{\mu\nu}\,^{\sigma\tau}
R_{\rho\lambda\sigma\tau}$ (where $\epsilon^{\mu\nu\rho\lambda}$
is a totally anti-symmetric tensor) and $\Delta (T,\bar{T})\propto
\ln \left[(T+\bar{T})|\eta(iT)|^4\right]$. The Dedekind
$\eta$-function is given by $\eta(iT)\equiv \e^{-\pi T/12}
\prod_{n\ge 1} \left(1-\e^{-2n\pi T}\right)$. There can be
additional terms in the four-dimensional effective Lagrangian,
such as,
\begin{equation}
{\cal L}_{\rm add} =  - V(S,T)-\cdots
\end{equation}
which includes, within the context of string theory, some
supersymmetry breaking non perturbative potentials coming from the
dynamics of branes, fluxes and orientifold planes, as well as the
back reaction effects from the localized sources. The potential
usually consists of sum of exponential terms determined by the
fluxes and the curvature terms~\cite{Ish:PRL1,Panda:2007}; this is
related to the fact that upon dimensional reduction of a gravity
theory, the potential is exponential in terms of canonically
normalized scalar fields descending from the internal space metric
and other modes.

To evaluate field equations obtained by varying a gravitational
action, we consider approximately homogeneous and isotropic
solutions given by the Friedmann-Robertson-Walker metric: $ds^2 =
-dt^2 + a^2(t)\, d {\bf x}^2$, where $a(t)$ is the scale factor of
the universe. $H\equiv \dot{a}/a$ defines the Hubble parameter and
the dot denotes a derivative with respect to cosmic time $t$.

In a flat FRW background, the terms proportional to $R\tilde{R}$
give a trivial contribution. Defining ${\rm Re} S=
e^{\varphi}/g_s^2$, ${\rm Re} T= e^{2\sigma}$, ${\rm Im} S\equiv
\tau= {\rm const}$ and ${\rm Im} T=0$, the four-dimensional
effective Lagrangian may be given by~\cite{Antoniadis:1992,ipn06b}
\begin{eqnarray}\label{dilatonGB2}
{\cal L}_{\rm eff} = \frac{R}{2\kappa^2} -\frac{\zeta}{2}
(\nabla\varphi)^2 -\frac{\gamma}{2} (\nabla\sigma)^2 + \frac{1}{8}
\left[\lambda f(\varphi)- \delta\, \xi(\sigma)\right] {\cal
R}^2-V(\varphi,\sigma),\end{eqnarray} where $\lambda\propto
1/g_s^2$, $g_s$ is four-dimensional string coupling, $\tau$ is
pseudoscalar axion and $\delta$ is a numerical constant. To
leading order in string loop expansion, $f(\varphi)\propto
\e^{\varphi}$ and $\xi(\sigma)=\ln 2-
\frac{\pi}{3}\,\e^{\sigma}+\sigma+4\sum_{n=1}^\infty \ln
(1-\e^{-2n\pi \e^{\sigma}})$. The latter implies that
$d\xi/d\sigma\simeq -{\rm sgn}(\sigma)
\frac{2\pi}{3}\sinh(\sigma)<0$. Several authors have explored
special features of the string-derived Lagrangian that might
provide some characteristic features of the above model (see for
example
~\cite{ipn06b,Kaloper,Easther,KRama,Kawai,Kanti,Easson,Tsujikawa,Ezawa,Piao}).

In the discussion below we consider the simplest case of a single
modulus, under the assumption that the compacification modulus
$\sigma$ would rapidly evolve along an instantaneous minimum
determined by the condition $dV/d\sigma= 0$, such that
$V(\varphi,\sigma) \approx {\rm const} \times V(\varphi)$, while
$\varphi$ attains a constant value only at late
times~\footnote{This assumption may just be reversed and assume
that the dilaton $\varphi$ would evolve more rapidly as compared
to $\sigma$; these all depend on an underlying model. Of course,
the single field description in terms of $\varphi$ (or $\sigma$)
could underestimate the actual evolution of the universe at early
epochs, like during inflation, because string compactifications
invariably involve more than one scalar field, and the
four-dimensional potential depends, in general, on all the moduli
field of the compactification. Nevertheless, this simple
approximation in the string-derived Lagrangian holds some validity
as a post-inflation scenario.}.

\section{Modified Gauss-Bonnet theory}

As should be clear from the above discussion, the simplest version
of scalar-tensor theories, which is perhaps sufficiently general
for explaining the present evolution of our
universe~\cite{Nojiri:2005}, may be given by~\cite{Carter:2005}
\begin{equation}\label{GB-action-II}
{\cal S}_{\rm eff} =\int d^4 x \sqrt{-g} \left[
\frac{R}{2\kappa^2}-\frac{\zeta}{2}(\nabla\varphi)^2 -V(\varphi)
+\frac{\lambda}{8} f(\varphi) {\cal R}^2 \right].
\end{equation}
Here $f(\varphi)$ is a function that, although computable in
concrete string models, may be taken to be general for the present
purpose. The most desirable property of the above type
modification of Einstein's theory is that only the terms which are
the second derivatives of the metric (or their product) appear in
field equations -- a feature perhaps most important in order to
make a gravitational theory absence of (spin-$2$)
ghosts~\cite{Ish:2001a} -- thereby ensuring the uniqueness of
their solutions~\cite{Dadhich:2005}. Of course, one can supplement
the above action with other higher derivative terms, such as those
proportional to $(\nabla_\mu\varphi \nabla^\mu\varphi)^2$ and
higher powers in $R$, $R_{\mu\nu}$ and
$R_{\mu\nu\rho\lambda}$~\cite{Maeda:2004,Sami:2005}, but in such
cases it would only be possible to get special (asymptotic)
solutions, so we limit ourselves to the above action.

Another important direction, which I will not review here, is the
quest for a concrete construction of four-dimensional cosmology
starting from some five-dimensional Gauss-Bonnet brane world
models~\cite{Mavromatos:2000,Neupane:2000,Lidsey:2003}. Of course,
in spacetime dimensions $D\ge 5$, a pure Gauss-Bonnet term can
lead to modification of Einstein field equation, even if
$f(\varphi)={\rm const}$, and hence influence the four-dimensional
cosmology defined on the $3$-brane. Here we limit ourselves to the
four-dimensional action and demand that $f(\varphi)$ is dynamical.
In this case the GB term ${\cal R}^2$ is not topological, rather
it can have an interesting dynamics, especially, on largest
cosmological scales.

\section{Cosmological perturbations and stability conditions }

To explore the stability of an effective gravitational action,
under large cosmological perturbations, one may consider the
following perturbed metric about a flat Friedmann-Robertson-Walker
(FRW) background:
\begin{equation}
ds^2= - (1+2\varpi)dt^2+ 2a \partial_i\chi dx^i dt +
a^2\left[(1+2\psi)\delta_{ij} + 2\partial_{ij} \eta +2
h_{ij}\right]dx^i\,dx^j,
\end{equation}
where $\varpi, \chi, \psi$, $\eta$ denote scalar and $h_{ij}$
denotes vector components of metric fluctuations, and
$\partial_{ij}\equiv \Delta_i\Delta_j- (1/3)\delta_{ij} \Delta^2$.
A remarkable property of the Gauss-Bonnet gravity is that the
linearised action can be expressed (in the absence of matter
fields) in the following explicit form~\cite{Hwang&Noh97}:
\begin{equation}
\delta^{(2)}{\cal S} \propto \int dt a^3 \left[-A(t) {\cal R}
\ddot{\cal R}+ \frac{B(t)}{a^2} \,{\cal R} \Delta^2{\cal
R}\right],
\end{equation}
where ${\cal R}$ is a gauge invariant quantity:
\begin{equation} {\cal R} \equiv \psi
-\frac{H}{\dot{\varphi}}\,\delta\varphi,
\end{equation} so-called a comoving perturbation.
For a linearized theory to be free of ghost and superluminal
modes, the following conditions
\begin{equation}
A(t)>0, \quad B(t) > 0,
\end{equation}
known as stability conditions, should perhaps be satisfied. For
quantum stability of (inflationary) solutions the speeds of
propagation of scalar and tensor modes should also remain
non-superluminal:
\begin{eqnarray}
0< c_{\cal R}^2 &=&
1+\frac{\mu^2\left[4\epsilon(1-\mu)-\lambda\kappa^2\left(
\ddot{f}-\dot{f} H\right)\right]}
{(1-\mu)\left(2\zeta(1-\mu)\varphi^{\prime\,2}+3\mu^2\right)}\le 1,\\
0< c_T^2 &=& \frac{1+\lambda\kappa^2\ddot{f}}{1+\lambda\kappa^2
\dot{f}H} =\frac{1-\mu^\prime+\epsilon\mu}{1-\mu}\le 1,
\end{eqnarray}
where $\epsilon\equiv \dot{H}/H^2=H^\prime/H$, $^\prime\equiv
d/d\ln{a}$ and $\mu\equiv - \lambda\kappa^2 \dot{f} H =\Omega_f$.
In fact, $A(t),\, B(t)<0$ implies a violation of unitarity, while
$B(t)> A(t)$ implies the existence of a superluminal propagation
or an ill-defined Cauchy problem. Moreover, in the case
$A(t)B(t)<0$, the system of equations could exhibit an exponential
type of instability, leading to an imaginary $c\Z{T}$. Below we
use the above relations for studying the stability of inflationary
solutions.

\section{Inflationary constraints}

One may put constrains on the strength of the coupling
$f(\varphi)$ by considering observational limits on the spectral
indices of scalar and tensor perturbations. For the present
theory, and in the limit that $c\Z{\cal R}^2, c\Z{T}^2 \approx
{\rm const}$, the spectral indices $n\Z{\cal R}$ and $n\Z{T}$ are
approximated by~\cite{Hwang&Noh97,Ohta07,Leith:2007}
\begin{equation}
n\Z{\cal R}-1=3-\Big\arrowvert \frac{3+\epsilon_1+2\epsilon_2+2
\epsilon_3}{1-\epsilon_1}\Big\arrowvert, \quad n\Z{T}=3-
\Big\arrowvert \frac{3-\epsilon_1+2
\epsilon_4}{1-\epsilon_1}\Big\arrowvert. \label{spectral-indices}
\end{equation}
where $\epsilon_1 \equiv -\frac{\dot{H}}{H^2}=-\epsilon$,
$\epsilon_2 = \frac{\ddot{\varphi}}{\dot{\varphi}H}$, $\epsilon_3
= \frac{\theta^\prime}{2\theta}$, $\epsilon_4 \equiv
-\frac{\mu^\prime}{2(1-\mu)}$, $\theta \equiv
\zeta+\frac{3\mu^2}{2(1-\mu){\varphi^\prime}^2}$ and $\mu \equiv
-\lambda \kappa^2 \dot{f} H$. One more quantity of cosmological
relevance is the tensor-to-scalar ratio, which, in the limit
$|\epsilon_1|\ll 1$, is approximated by
\begin{equation}
r \approx 16 \frac{2\zeta x^2(1-\mu)+3\mu^2}{(2-\mu)^2}
\left(\frac{c_{\cal R}}{c_T}\right)^{3}.
\end{equation}
WMAP data alone puts the constraints $0.94 <n_{\cal R}<0.98$ and $r<0.28$ for a
single scalar field model.

\section{Non-singular inflationary solutions}

\subsection{Absence of scalar potential}

 Let us consider a non-singular inflationary solution obtainable by dropping
the scalar potential. To quantify this, one sets $V(\varphi)=0$.
One also defines
\begin{equation}\label{def-for-nu}
{\cal F}(\varphi) \equiv - \lambda f(\varphi) H^2.
\end{equation}
The magnitude of ${\cal F}$ should decrease with the expansion of
the universe, so that all higher-order corrections to Einstein's
theory become only sub-leading~\footnote{Particularly in the case
${\cal F}\simeq {\rm  const}\equiv {\cal F}\Z{0}$, the coupled
term $\lambda f(\varphi) {\cal R}^2$ is subleading to the Einstein
term $R/2\kappa^2=3 M_P^2(2H^2+\dot{H})$ for $|{\cal F}\Z{0}| \ll
1$ or $\lambda\ll 1$.}. With ${\cal F} \equiv {\cal F}\Z{0}$, the
explicit solution is given by
\begin{equation}\label{soln-epsilon}
\frac{\dot{H}}{H^2}=- {\cal A} + {\cal B} \tanh {\cal B} (N+C),
\end{equation}
where $N\equiv \ln{a}$, $C$ is an integration constant and
\begin{equation} {\cal A}
\equiv \frac{5 {\cal F}\Z{0} +1}{2 {\cal F}\Z{0}}, \quad {\cal B}
\equiv \sqrt{{\cal A}^2-6{\cal A}+15}
\end{equation}
The Hubble parameter is $H\propto \e^{-{\cal A} N} \cosh {\cal B}
(N+C)$. The ${\cal F}\Z{0} >0$ solution, which allows $\dot{H}>0$,
supports a super-luminal expansion, see also \cite{Kanno:2006}. It
is possible to get a red-tilted scalar index ($n_{\cal R}<1$) for
${\cal F}\Z{0}> - 2/3$.

\subsection{Inflating with an exponential potential}

\begin{figure}[ht]
\begin{center}
\hskip-0.3cm
\epsfig{figure=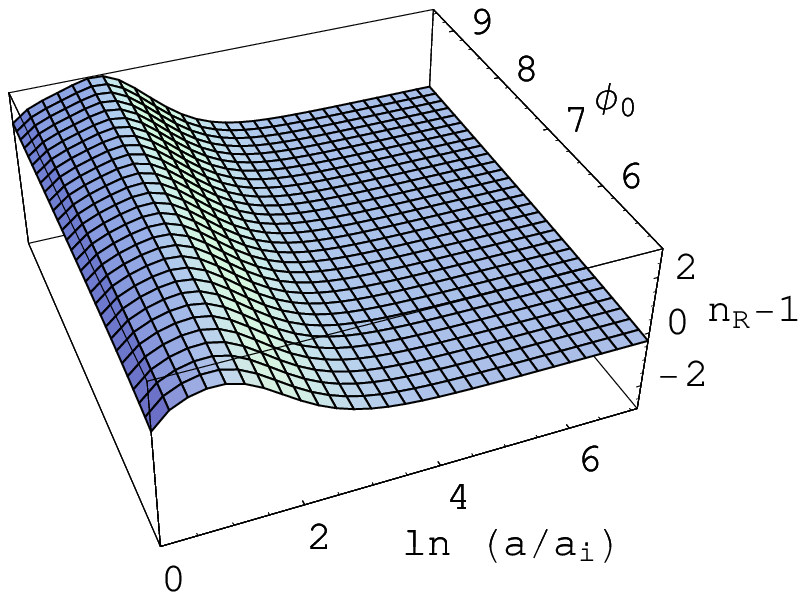,height=2.3in,width=2.5in}
\hskip-0.2cm
\epsfig{figure=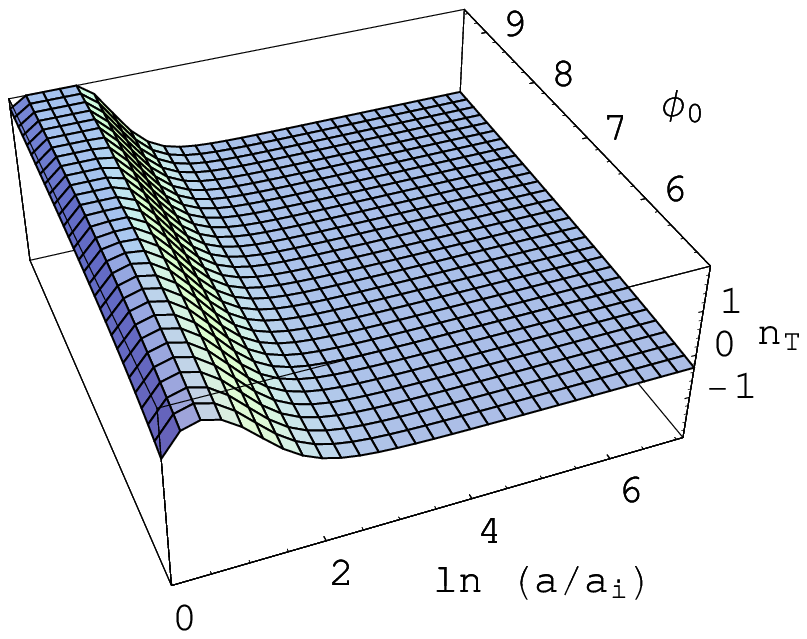,height=2.3in,width=2.5in}
\end{center}
\caption{The spectral indices $n\Z{R}$ and $n\Z{T}$ as the
functions of $\phi\Z{0}= 2\sqrt{\zeta}\varphi\Z{0}/\beta$ and $\ln
(a/a\Z{i})$.}\label{Fig1}
\end{figure}

Consider that $V(\varphi)\propto e^{-\beta (\varphi/\varphi_0)}$
and $f_{,\varphi} H^2\propto \varphi^\prime$; the latter choice is
motivated by the fact that the coupling takes the form
$f(\varphi)\propto e^{\beta(\varphi/\varphi_0)}$ in the limit
$\varphi^\prime\to {\rm const}$, or after a few e-folds of
inflation. The explicit solution is
\begin{equation}\label{expo-exact-sol}
\varphi=\frac{2}{\beta}\varphi\Z{0}
\ln\frac{a/a_i}{\cosh\chi\ln(a/a_i)}+{\rm const},\quad
\frac{\dot{H}}{H^2}
=\frac{2\zeta{\varphi^\prime}^2}{6+\zeta{\varphi^\prime}^2}
-\frac{\beta}{\varphi\Z{0}}\,{\varphi^\prime}^2,
\end{equation}
where $\chi\equiv
\sqrt{(2\zeta\varphi\Z{0}^2-3\beta^2)/2\zeta\varphi\Z{0}^2}$,
$\varphi^\prime\equiv d\varphi/d(\ln a)=\dot{\varphi}/H$ and $a_i$
is the initial value of scale factor $a(t)$. From this solution we
can easily evaluate the indices $n\Z{\cal R}$ and $n\Z{T}$,
using~(\ref{spectral-indices}). The observation in Fig.~\ref{Fig1}
that $n\Z{R}\simeq 3$ at some early stage of inflation is not
quite correct since in that region $c_{\cal R}^2$ and $c_T^2$ are
varying considerably, for which there would be non-trivial
corrections to the formulae (\ref{spectral-indices}). A result
consistent with the WMAP data (e.g. $n\Z{R} \simeq 0.96$ and
$n\Z{T} <0.2$) can be obtained for $|\beta/\varphi\Z{0}| <
\sqrt{\zeta}/4$.

\section{Matter-scalar couplings}

For constructing a realistic late time cosmology, one should
consider the ordinary fields (matter and radiation) and also their
natural interactions with the scalar field $\varphi$.

\subsection{Minimally coupled scalar field}

In a flat FRW spacetime, the Gauss-Bonnet term ${\cal R}^2$
vanishes only at the stage of zero acceleration, and it flips its
sign once the universe begins to accelerate. This effect can
overturn the slope of the effective potential:
$$\Lambda(\varphi)\equiv V(\varphi) - \frac{\lambda}{8} f(\varphi)
{\cal R}^2, $$ and push the universe transiently to a phantom era.
Such an effect can be seen by considering the effective equation
of state:
\begin{equation}
w_{\rm eff}\equiv -1-\frac{2\dot{H}}{3H^2}= \frac{p_{\rm
tot}}{\rho_{\rm tot}} =w_m \Omega_m+ w_r \Omega_r + w_{\varphi}
\Omega_{\varphi},
\end{equation}
where $m={\rm matter}$ and $r={\rm radiation}$. With the
assumption that the ordinary matter is approximated by a
non-relativistic perfect fluid (i.e. $w_m\simeq 0$ and
$\Omega_r\ll 1$), we find $w_{\rm eff}\simeq w_{\varphi}
\Omega_{\varphi}$. A simple calculation shows
$\rho_{\varphi}+p_{\varphi}=\zeta\dot{\varphi}^2+ \lambda
H^2(\ddot{f}- \dot{f} H) +2 \lambda H \dot{H}
\dot{f}$~\cite{Carter:2005}. To this relation, the stability
conditions $1> \kappa^2 |\lambda \ddot{f}|$ and $|\ddot{f}|\ge
|\dot{f}H|$ may be imposed, so as to keep the propagation speed of
tensor and scalar modes non-superluminal. Nevertheless, it is
possible to get $p_{\varphi}+\rho_{\varphi} <0$, or
$w_\varphi<-1$, without making the cosmic expansion superluminal,
or violating the condition $\dot{H}\le 0$. This simple picture has
obvious and intuitive appeal.

\subsection{Non-minimally coupled scalar field}

The constraints on the modified Gauss-Bonnet gravity may arise by
two different dynamics: one is the standard interaction effect
between the scalar field $\varphi$ and the Gauss-Bonnet term,
while the other is the effect of nonminimal coupling between the
scalar field $\varphi$ and matter. The latter effect might perhaps
be more significant than the former, especially, while applying
the model into high density regions, or solar system experiments.
To this reason, let us write the matter Lagrangian in a general
form:
\begin{equation}
{\cal S}_{\rm matter}={\cal S}(A^2(\varphi)g_{\mu\nu}, \psi_m),
\end{equation}
where $A(\varphi)$ measures the response of the geometry due to a
time-variation of the field $\varphi$. Ordinary fields (matter and
radiation) couple to $A^2(\varphi) g_{\mu\nu}$ rather than the
Einstein metric $g_{\mu\nu}$ alone. Indeed, $\varphi$ couples to
the trace of the matter stress tensor, $g_{(i)}^{\mu\mu}
T_{\mu\nu}^{(i)}$, so the radiation term (for which $w_r=1/3$)
does not contribute to the (Klein-Gordon) equation of motion for
$\varphi$:
\begin{equation}\label{KG-evolution}
\dot{\rho}\Z{\varphi} + 3 H \rho\Z{\varphi} \left(1 +
w\Z{\varphi}\right) = -\dot{\varphi}(1-3 w_i) \alpha\Z{\varphi}
A(\varphi) \rho\Z{m},
\end{equation}
where $\rho\Z{\phi} \equiv \frac{\zeta}{2} \dot{\phi}^2 +
V(\phi)-3\lambda H^3 \dot{f}$, $w\Z{\phi}\equiv
p\Z{\phi}/\rho\Z{\phi}$ and $w\Z{i}\equiv p\Z{i}/\rho\Z{i}$. In
order for current experimental limits on verification of the
equivalence principle to be satisfied, the quantity
\begin{equation}
\alpha\Z{\varphi}\equiv \frac{d\ln
A(\varphi)}{d\,(\kappa\varphi)},
\end{equation}
which measures the coupling of $\varphi$ to background (baryonic
and dark) matter, must be much smaller than unity, at least, on
cosmological scales. The local GR constraints on $\alpha_\varphi$
and its derivatives imply that~\cite{EspositoFarese:2004}
\begin{equation}
\alpha\Z{\varphi}^2 \le 4\cdot 10^{-5}, \qquad
\beta\Z{\varphi}=\frac{d\alpha\Z{\varphi}}{d\varphi} > -4.5.
\end{equation}
On large cosmological scales, where $\rho\Z{m}\lesssim
\rho\Z{\varphi}$ and $\kappa^2 V(\varphi)\sim 3 H\Z{0}^2$,
$\varphi$ is expected to be sufficiently light, as for {\it
quintessence}, $m_\varphi \equiv \sqrt{V_{\varphi\varphi}} \sim
{10}^{-33} {\rm eV}$, and the term on r.h.s. of
eq.~(\ref{KG-evolution}) may be safely ignored. In fact, in
ref.~\cite{Damour:02a}, smallness of $\alpha\Z{\varphi}$ was found
to be linked to the smallness of the horizon-scale cosmological
density fluctuation, $\delta \rho/\rho \sim 5\times {10}^{-5}$ (at
the surface of last scattering). However, in high density regions,
or within galactic distances, $\delta \rho/\rho \gg {10}^{-5}$ and
$\varphi$ can be massive, like $m_\varphi \gtrsim 10^{-3} {\rm
eV}$, in which case the observable deviations from Einstein's
gravity are normally quenched on distances larger than a fraction
of millimeter.

If $A(\varphi)$ is sufficiently flat near the current value of
$\varphi=\varphi_0$, then the matter-scalar coupling can have only
modest effects on cosmological scales. Especially, in the case
that $A(\varphi) \propto e^{Q (\varphi/M_P)}$, the above GR
constraints may be satisfied only for a small $Q$ ($\ll 1$). This
restriction on the slope (or strength) of matter-scalar coupling
may not apply to a gravitationally bound system, or in high
density regions, where the field $\varphi$ is not essentially
light or weakly coupled to matter degrees of freedom (of the
standard model). The latter argument is actually consistent with
ideas widely used in recent experiments aimed to detect axion-like
particles.

\section{Late-time cosmology}

Making just one simplifying assumption that $\varphi \equiv
\varphi_0\, \ln [a(t)] +{\rm const}$, and then inverting the field
equations following from ({\ref{GB-action-II}), we
find~\cite{ipn06b}
\begin{equation}
f(\varphi)= - f_0\, \e^{\beta (\varphi/\varphi_0)}- f_1, \quad
V(\varphi)= \frac{2(\delta-1)}{3\lambda\,\kappa^4}
\left(\frac{df(\varphi)}{d\varphi}\right)^{-1}\equiv
V_0\,\e^{-\beta(\varphi/\varphi_0)},
\end{equation}
where $\beta=1+3\delta$ and $\delta= \kappa^2\varphi_0^2/2$. These
simplest choices for the potential and the scalar coupling admit
the following simple solution~\cite{Leith:2007}
\begin{equation}
a(t)=a_0\,t^{2/\beta},
\end{equation}
satisfying the relations: \begin{equation} {\lambda f_0}
=\frac{(\beta-2\zeta\delta) \beta}{2\kappa^2(\beta+2)}, \quad V_0=
\frac{24(2-\beta)+8\zeta\delta(10-\beta)}{(\beta+2)\beta^2
\kappa^2}.
\end{equation}
Acceleration requires $\beta < 2$ (if it is to be future eternal);
thus, for the model to provide a solution to the dark energy
problem, the strength of the GB coupling must grow with time,
$f(\varphi)\propto \e^{\beta (\varphi/\varphi_0)}\propto
{a(t)}^\beta$. This is actually consistent with superstring models
studied in~\cite{Antoniadis:1992,Kaloper,Carter:2005}. One should,
however, note that a growing $f(\varphi)$ does not necessarily
mean that the term $f{\cal R}^2$ will dominate at late times the
potential and/or the Einstein-Hilbert term. In the present
universe $H_0\sim 10^{-60} M_P$, which leads to $\kappa^2 V \simeq
10^{-120} M_P^2$, $R/6 \simeq H_0^2\sim 10^{-120} M_P^2$ and
$\kappa^2 f {\cal R}^2 \propto V^{-1} H_0^4 \equiv f_0\,e^{-120}
M_P^2$. For $f_0\ll 1$, $f {\cal R}^2$ is only subleading to $V$
and $R/\kappa^2$. Terms higher powers in Ricci scalar ($R^n$ with
$n\ge 3$) contribute with $H_0^{2n}$ and are thus subleading to
$f(\varphi) {\cal R}^2$.

\begin{figure}[ht]
\begin{center}
\hskip-0.2cm
\epsfig{figure=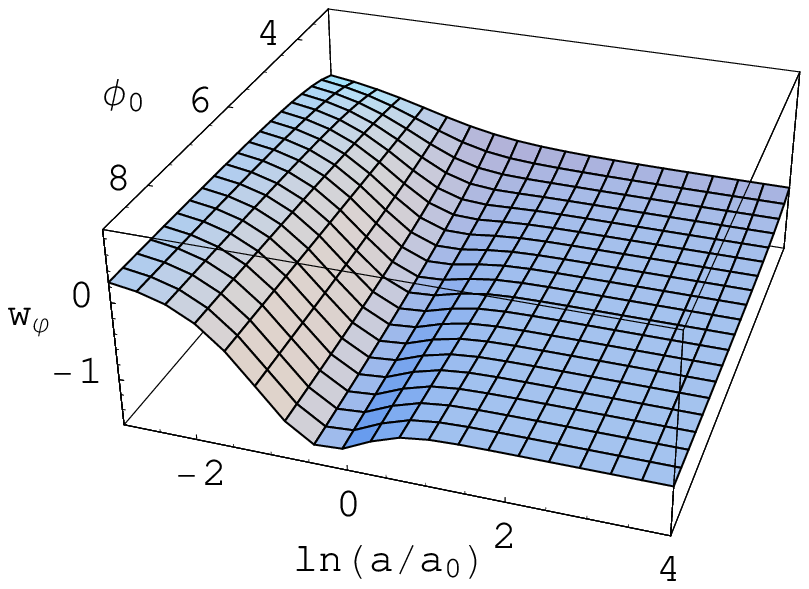,height=2.0in,width=2.3in}
\hskip-0.3cm
\epsfig{figure=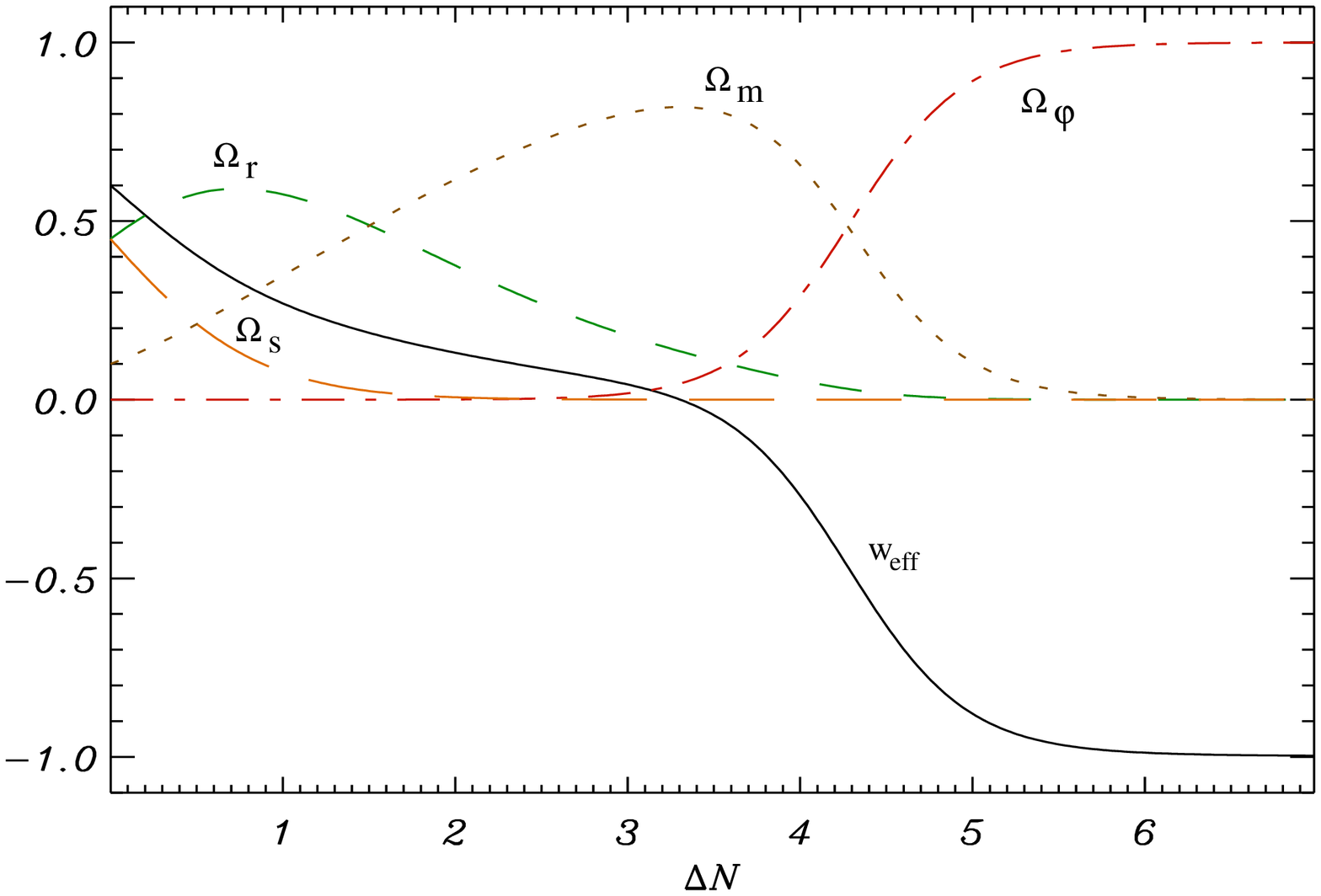,height=2.2in,width=2.7in}
\end{center}
\caption{(Left plot) The dark energy equation of state as a
function of $\phi\Z{0}$ and $\ln (a/a\Z{0})$. (Right plot) The
evolution of the fractional densities: $\Omega\Z{\rm m}$ (dots,
brown), $\Omega\Z{\rm r}$ (dashes, green), $\Omega\Z{\rm s}$ (long
dashes, orange), $\Omega\Z{\varphi}$ (dot-dash, red) and
$\Omega\Z{\rm GB}\equiv \Omega\Z{f}$ (dot-dot-dot-dash, blue) and
the effective equation of state $w\Z{\rm eff}\equiv
-1-2\dot{H}/3H^2$, with $\zeta=1$, $(\beta/\varphi\Z{0})=3$,
$\alpha\Z{\varphi}^2 = 10^{-5}$ (dust) and $\alpha\Z{\varphi}^2 =
10^{-2}$ (stiff matter). The initial values are $\Omega\Z{f}^{\rm
i}= 10^{-6}$, $\varphi^\prime_{i} = 10^{-7}~M\Z{P}$,
$\left(V/H^2\right)\Z{\rm i}= 3\times 10^{-15}$. $\Delta N\equiv
\ln a+C$; here $C$ may be chosen such that $\ln a=0$ corresponds
to $\Omega\Z{m}\simeq 0.27$ and
$\Omega\Z{\varphi}+\Omega\Z{f}\simeq 0.73$.}\label{Fig2}
\end{figure}

Let us consider a specific model for which the dark energy
equation of state becomes less than $-1$, but only transiently.
This example is provided by the choice $V(\varphi)\propto
e^{-\beta(\varphi/\varphi\Z{0})}$ and $f_{,\varphi} H^2\propto
\varphi^\prime$. In this case the explicit solution is given by
(\ref{expo-exact-sol}). One may take $a\Z{i}= a\Z{0}\equiv 1$, so
that $a(t)< 1$ in the past. As shown on the left panel of
Fig.~\ref{Fig2}, the equation of state $w \equiv
-1-\frac{2\dot{H}}{3H^2}$ becomes less than $-1$, but only
transiently, for $\phi\Z{0}\equiv
2\sqrt{\zeta}\varphi\Z{0}/\beta\gtrsim 5$. This behaviour may be
seen also in the presence of matter field, see
ref.~\cite{Leith:2007} for details.

The right panel of Fig.~\ref{Fig2} represents a characteristic
evolution of the universe for which the Gauss-Bonnet term never
becomes dominant, or it contributes only sub-dominantly. In this
plot the coupling $f(\varphi)$ has been chosen such that
$\varphi^\prime f_{,\varphi} H^2\simeq {\rm const}$ and the
Gauss-Bonnet energy density fraction is (almost) constant,
$\Omega\Z{f}\sim 10^{-6}$. With such a small contribution of the
coupled GB term $\lambda f(\varphi) {\cal R}^2$, almost every
constraints on the model may be satisfied, including the BBN bound
($\Omega\Z{\varphi}(1 ~{\rm MeV})< 0.1$) and solar system
constraints (see below).

\begin{figure}[ht]
\begin{center}
\hskip-0.3cm \epsfig{figure=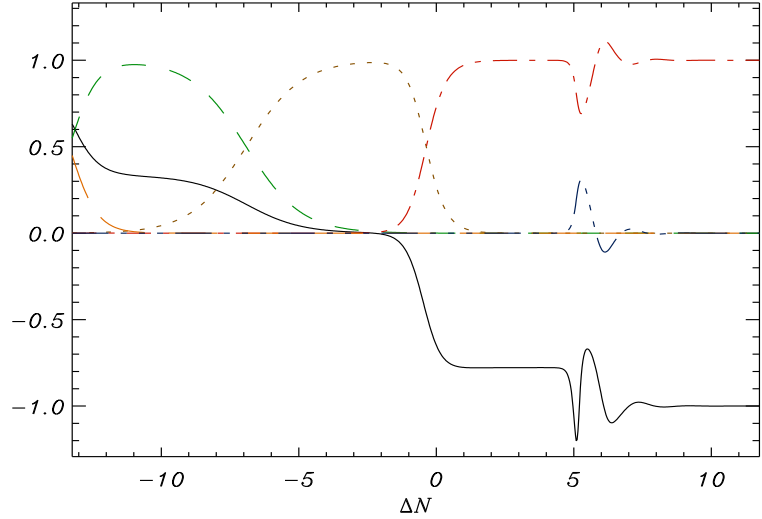,height=3in,width=4.5in}
\end{center}
\caption{As in the right panel of Fig.~\ref{Fig2}, but with the
parametrization $f(\varphi)\propto
e^{\alpha(\varphi/\varphi\Z{0})}$, and the choice
$\alpha=12\varphi\Z{0} \gg \beta =\sqrt{2/3} \varphi\Z{0}$. For
the large $\alpha$ it is not unnatural that the coupled
Gauss-Bonnet term $f(\varphi) {\cal R}^2$ becomes significant
(non-negligible) at recent times, or even at distant future. Here
$\Delta N\equiv \ln (a/a\Z{0})$, we normalize the scale factor
such that $\ln a=0$ at $a\Z{0}=1$. }\label{Fig3}
\end{figure}

We can construct an explicit model by using the parametrization
$f(\varphi)\equiv f_0\, e^{\alpha(\varphi/\varphi_0)}$ and
$V(\varphi) \equiv V_0\,
e^{-\beta(\varphi/\varphi_0)}$~\cite{Koivisto:2006a} and also
replicate many observable properties of the universe from
nucleosynthesis to the present
epoch~\cite{Koivisto:2006b,Leith:2007} (see
also~\cite{Nojiri:2006})). A possible drawback of this simple
parametrization is, however, that, especially, for large slope
parameters, like $\alpha> \beta \gtrsim \sqrt{3} (\varphi_0/M_P)$,
the model may exhibit some kind of semi-classical instabilities
associated with the linearized inhomogeneities or quantum
fluctuations that grow explosively as the limit $c_T^2 <0$ is
approached or the tensor modes start to propagate faster than
light's velocity~\cite{Calcagni:2006}.

This rather undesirable feature of the model is indeed related to
the fact that, for $\alpha \gg \beta$, the contribution of GB term
become appreciable (or non-negligible) at recent times (or even in
far future) but only transiently. In the case GB contribution
becomes appreciable, even momentarily, one normally observes an
oscillatory crossing of $w\Z{\varphi}= -1$. Generally, the
amplitude of these oscillations corresponds to the amplitude of
the oscillations seen in the Gauss-Bonnet contribution and hence
is heavily dependent on the slope of the scalar-GB coupling,
$\alpha$. For large $\alpha$ one may observe much larger
oscillations, which, however, disappear when the GB contribution
becomes negligibly small, and settle to a late time evolution for
which $w\Z{\varphi} \approx -1$. In most cases, this limit is
approached from above, so the issue inherent with a
super-inflation or a violation of unitarity may not be applicable
to late time cosmologies. At any rate, the appearance of a
superluminal mode, though not inevitable, could actually imply
that one would have to invoke modifications of the simplest
exponential parametrization or should allow only small slope
parameters.

\section{Time-variation of fundamental constants}

Scalar-tensor theories of gravity also entertain the result that
some of the fundamental constants of nature may vary with time,
including the Newton's constant, which are however tightly
constrained by observations. On large cosmological scales, it is
reasonable to assume that $A(\varphi)={\rm const}$. In this case,
the growth of matter fluctuations in the Gauss-Bonnet theory can
be expressed in the following standard form:
\begin{equation}
\ddot{\delta}+2\dot{\delta} H=4\pi {G}_* \rho_m \,\delta,
\end{equation}
where the normalized Newton's constant $G_*$ may be given
by~\cite{Amendola:2005}
\begin{equation}
{G}_* ={G} \left[1+3 \,\Omega_f -\frac{\dot{\varphi}}{H}\left(
\frac{\ddot{\varphi}}{\dot{\varphi}^2}+
\frac{f_{\varphi\varphi}}{f_\varphi}\right)\Omega_f \right],
\end{equation}
where $\Omega_f\equiv -\lambda \kappa^2 \dot{\varphi} H
f_\varphi$. Unlike the slow roll relations
$\ddot{\varphi}/\dot{\varphi}, \,\dot{\varphi} \ll 1$, the ratios
like $\ddot{\varphi}/\dot{\varphi}^2$ and $\ddot{f}/\dot{f}$,
which appear in the expression
\begin{equation}
 \frac{f_{\varphi\varphi}}{f_\varphi}
=\frac{d^2 f/d\varphi^2}{df/d\varphi} =\frac{H}{\dot{\varphi}}
\left(\frac{\ddot{f}}{\dot{\varphi}
\dot{f}}-\frac{\ddot{\varphi}}{\dot{\varphi}^2}\right)
\end{equation}
can be of order unity (in units $M_P=1$). It is not improbable
that ${G}_* \approx G$ for present value of the field,
$\varphi_0$, and the coupling, $f(\varphi_0)$. In fact, almost
every models of scalar-tensor gravity behave as Einstein's GR
supplemented with a cosmological constant term $\Lambda$, if $
\varphi^\prime= \frac{\dot{\varphi}}{H} \ll M\Z{P}$ holds (at
least) after the epoch of big bang nucleosynthesis. In the
particular case that $f(\varphi)\propto
e^{\beta(\varphi/\varphi_0)}$, we obtain
\begin{eqnarray}
{G}_*
&=& G \left[1+ \lambda f(\varphi) H^2
\left(\varphi^{\prime\prime}+\epsilon
\varphi^\prime-2\varphi^\prime \right)\right].
\end{eqnarray}
Thus one should satisfy, at least, one of the following
conditions: (i) $|\lambda|\ll 1$, (ii) $|f(\varphi)|H^2\ll 1$, or
(iii) $|\varphi^\prime|= |\dot{\varphi}/H|\ll M\Z{P}$, in order to
get $G_*\simeq G$ at present. For a specific model studied in
\cite{Damour:02a}, a safe upper bound is found to be
$|\varphi^\prime_0|< 0.84\,M\Z{P}$. Nevertheless, within solar
system and laboratories distance, there exists a more stronger
bound that $(d{G}_*/dt)/G_* < 0.01\, H_0$ (where $H_0$ is the
Hubble expansion rate at present). This last condition translates
to the constraint $|G_{\rm now}-G_{\rm nucleo}|/G_{\rm now}(t_{\rm
now}-t_{\rm nucleo}) < 10^{-12} {\rm yr}^{-1}$. The quantity
$d{G}_*/dt$ is actually suppressed (as compared to $G_*$) by a
factor of $\dot{\varphi}/H$, so it is necessary to satisfy
$\varphi^\prime\ll M\Z{P}$, at least, on large cosmological
scales. Another opportunity for the model to overcome local
gravity constraints coming from GR is to have a coupling
$f(\varphi_0)$ which is nearly at its minimum. This is very much
the approach one takes in a standard scalar-tensor theory.

\section{Further constraints}

The growth of matter perturbations and the integrated Sachs-Wolfe
(ISW) effect are the other effective ways of constraining the
model under consideration~\cite{Amendola:2005}. In the case
$\lambda f(\varphi) {\cal R}^2$ is subdominant to $V(\varphi)$
(thus $|\Omega_f| \ll 1$), the matter growth factor may be
approximated by
\begin{equation}
\left(\frac{\dot{\delta}}{\delta}\right)_{\rm EGB} \approx
\left(\frac{\dot{\delta}}{\delta}\right)\left[1-
\left(1+\frac{H^\prime}{H}\right)\left(1+0.75\, \Omega_m\right)
\,\Omega_f \right].
\end{equation}
In an accelerating spacetime, so $H^\prime/H \ge -1$, the
Gauss-Bonnet coupling decreases the matter growth factor (as
compared to the standard $\Lambda$CDM), for $\Omega_f>0$. In view
of the observational uncertainly in the growth rate of large scale
structures~\cite{SDSS}, $f_{\rm struc} \equiv
\left({\dot{\delta}}/{\delta}\right) =0.51\pm 0.1$, the
Gauss-Bonnet energy density fraction $\Omega_f$ should perhaps not
exceed $20\%$~\cite{Amendola:2005}. This last result may apply
only to largest cosmological scales, and it is, by no means,
applicable to gravitationally bound systems, such as, our solar
system.

Under the post-Newtonian approximation:
\begin{equation}
ds^2=-(1+2\Phi) (c dt)^2 + (1-2\Psi) \delta_{ij} dx^i\,dx^j
\end{equation}
where $\Phi, \Psi \sim {\cal O} (G M/r c^2)$, the solar system
constraints appear to be stronger than astrophysical constraints,
mainly, due to a small fractional anisotropic
stress~\cite{Amendola:2007} (see also~\cite{Sotiriou:2006}):
\begin{equation}
\hat{\gamma}-1\equiv \frac{\Psi-\Phi}{\Phi} \approx 2\Omega_{f}
\left(1-\frac{\dot{\varphi}}{H}
\left(\frac{\ddot{\varphi}}{\dot{\varphi}^2}
+\frac{f_{\varphi\varphi}}{f_\varphi}\right)\right) < 4\times
10^{-5}.
\end{equation}
When applied to solar system distances, the above expression
demands that $\Omega\Z{f} \lesssim 10^{-5}$. There remains the
possibility that the classical tests of Newtonian gravity, which
typically deal with small perturbations in fixed (or
time-independent) backgrounds are almost unaffected by the GB type
modification of Einstein's theory.

\section{Conclusions}

The important ingredient of the present approach to dark energy
cosmology is the treatment of gravitational coupling between the
dynamical scalar field $\varphi$ and the quadratic Riemann
invariant of the Gauss-Bonnet form, which gives rise to
nonsingular cosmologies for a wide range of scalar-curvature
couplings. The model also provides plausible explanation to some
outstanding cosmological conundrums, including: the transition
from matter dominance to dark energy and the late time cosmic
acceleration. Furthermore, the scalar-curvature coupling can
easily trigger onset of a late dark energy domination. Despite
these promising signs, it remains to be checked whether the
Gauss-Bonnet modification of Einstein's theory will lead to
genuine contact between observations and string theory.

String theory is known to be free from ghosts and superluminal
modes, at least, in a flat ten-dimensional Minkowski background.
This is perhaps not essentially the case in a four-dimensional FRW
background. The effective string actions in four dimensions may
well exhibit some unwarranted features, such as, short scale
instabilities and superluminal propagation of tensor or scalar
modes, under inhomogeneous (cosmological) perturbations. The model
discussed here is perhaps not an exception; at least, for the
potential and the Gauss-Bonnet coupling in simplest exponential
forms, one could see that the tensor or scalar modes propagate at
a speed faster than light at some stage, especially, for large
slope parameters. The appearance of a superluminal mode is
generic, and perhaps also acceptable, if such an effect is only
transient.

In the present proposal for explaining a crossing of cosmological
(dark energy) equation of state, $w\Z{DE}=-1$, and a superluminal
propagation of scalar or tensor modes, a number of important,
physically falsifiable predictions can be made. These include a
transient violation of Lorentz symmetry and the weak equivalence
principle, associated with the microscopic effects of the coupling
between $\varphi$ and background matter in high density regions.

Ninety years after Einstein's proposition of general relativity
with a cosmological constant, a modified cosmological scenario
with its natural generalization is close to experimental test and
possibly an outlet compatible with present experimental data. The
coming generation of cosmological experiments, including Dark
Energy Survey, will probably rule out the great majority of
string-derived models, as well as exclude those class of
scalar-tensor theories which give rise to unphysical states. {\sl
Time will tell}.

\medskip

{\it Acknowledgements}: This research is supported by the
Foundation for Research, Science and Technology (New Zealand)
under Research Grant No. E5229 and also by Elizabeth Ellen Dalton
Research Award (E5393).




\begin{thebibliography}{99}
\itemsep 0pt

\bibitem{Supernova}
S. Perlmutter {\it et al.} [Supernova Cosmology Project
Collaboration], Astrophys. J. {\bf 517}, 565 (1999); A.~G. Riess,
{\it et al.} [Supernove Search Team Collaboration], Astrophys. J.
{\bf 560}, 49 (2001).

\bibitem{WMAP03}
D.~N. Spergel {\it et al.} [WMAP Collaboration], Astrophys. J.
Suppl. {\bf 148}, 175 (2003).

\bibitem{Veneziano:2000}
  G.~Veneziano,
  arXiv:hep-th/0002094.

\bibitem{Antoniadis:1992}
  I.~Antoniadis, E.~Gava and K.~S.~Narain,
  Nucl.\ Phys.\  B {\bf 383}, 93 (1992) [arXiv:hep-th/9204030];
  I.~Antoniadis, J.~Rizos and K.~Tamvakis,
  Nucl.\ Phys.\  B {\bf 415}, 497 (1994) [arXiv:hep-th/9305025].

\bibitem{Ish:PRL1}
I.~P.~Neupane,
  Phys.\ Rev.\ Lett.\  {\bf 98}, 061301 (2007) [arXiv:hep-th/0609086].

\bibitem{Panda:2007}
  S.~Panda, M.~Sami and S.~Tsujikawa,
  Phys.\ Rev.\  D {\bf 76}, 103512 (2007) [arXiv:0707.2848].

\bibitem{ipn06b}
  I.~P.~Neupane,
  Class.\ Quant.\ Grav.\  {\bf 23}, 7493 (2006)  [arXiv:hep-th/0602097];
  arXiv:hep-th/0605265.

\bibitem{Kaloper}
  N.~Kaloper, R.~Madden and K.~A.~Olive,
  Nucl.\ Phys.\  B {\bf 452}, 677 (1995) [arXiv:hep-th/9506027].

\bibitem{Easther}
  R.~Easther and K.~i.~Maeda,
  Phys.\ Rev.\  D {\bf 54}, 7252 (1996) [arXiv:hep-th/9605173].

\bibitem{KRama}
  S.~Kalyana Rama,
  Phys.\ Lett.\  B {\bf 408}, 91 (1997) [arXiv:hep-th/9701154].

\bibitem{Kawai}
  S.~Kawai, M.~a.~Sakagami and J.~Soda,
  Phys.\ Lett.\  B {\bf 437}, 284 (1998) [arXiv:gr-qc/9802033];
  S.~Kawai and J.~Soda,
  Phys.\ Lett.\  B {\bf 460}, 41 (1999) [arXiv:gr-qc/9903017].

\bibitem{Kanti}
  P.~Kanti, J.~Rizos and K.~Tamvakis,
  Phys.\ Rev.\  D {\bf 59}, 083512 (1999) [arXiv:gr-qc/9806085].

\bibitem{Easson}
  D.~A.~Easson and R.~H.~Brandenberger,
  JHEP {\bf 9909}, 003 (1999) [arXiv:hep-th/9905175].

\bibitem{Tsujikawa}
  S.~Tsujikawa,
  Phys.\ Lett.\  B {\bf 526}, 179 (2002) [arXiv:gr-qc/0110124].

\bibitem{Ezawa}
  Y.~Ezawa, H.~Iwasaki, M.~Ohmori, S.~Ueda, N.~Yamada and T.~Yano,
  Class.\ Quant.\ Grav.\  {\bf 20}, 4933 (2003) [arXiv:gr-qc/0306065].

\bibitem{Piao}
  Y.~S.~Piao, B.~Feng and X.~m.~Zhang,
  Phys.\ Rev.\  D {\bf 69}, 103520 (2004) [arXiv:hep-th/0310206].

\bibitem{Nojiri:2005}
  S.~Nojiri, S.~D.~Odintsov and M.~Sasaki,
  Phys.\ Rev.\  D {\bf 71}, 123509 (2005) [arXiv:hep-th/0504052].

\bibitem{Carter:2005}
  B.~M.~N.~Carter and I.~P.~Neupane,
  Phys.\ Lett.\  B {\bf 638}, 94 (2006) [arXiv:hep-th/0510109];
   B.~M.~N.~Carter and I.~P.~Neupane,
  JCAP {\bf 0606}, 004 (2006) [arXiv:hep-th/0512262].

\bibitem{Ish:2001a}
  Y.~M.~Cho, I.~P.~Neupane and P.~S.~Wesson,
  Nucl.\ Phys.\  B {\bf 621}, 388 (2002) [arXiv:hep-th/0104227].

\bibitem{Dadhich:2005}
  N.~Dadhich,
  arXiv:hep-th/0509126.

\bibitem{Maeda:2004}
  K.~i.~Maeda and N.~Ohta,
  Phys.\ Rev.\  D {\bf 71}, 063520 (2005) [arXiv:hep-th/0411093].

\bibitem{Sami:2005}
  M.~Sami, A.~Toporensky, P.~V.~Tretjakov and S.~Tsujikawa,
  Phys.\ Lett.\  B {\bf 619}, 193 (2005) [arXiv:hep-th/0504154];
  G.~Calcagni, S.~Tsujikawa and M.~Sami,
  Class.\ Quant.\ Grav.\  {\bf 22}, 3977 (2005) [arXiv:hep-th/0505193];
  E.~Elizalde, S.~Jhingan, S.~Nojiri, S.~D.~Odintsov, M.~Sami and I.~Thongkool,
  arXiv:0705.1211;
  K.~Bamba, Z.~K.~Guo and N.~Ohta,
  arXiv:0707.4334.

\bibitem{Mavromatos:2000}
  N.~E.~Mavromatos and J.~Rizos,
  Phys.\ Rev.\  D {\bf 62}, 124004 (2000) [arXiv:hep-th/0008074];\\
I.~P.~Neupane,
  JHEP {\bf 0009}, 040 (2000) [arXiv:hep-th/0008190].

\bibitem{Neupane:2000}
  I. P. Neupane  
  Phys.\ Lett.\  B {\bf 512}, 137 (2001) [arXiv:hep-th/0104226];
  I. P. Neupane,
  Class.\ Quant.\ Grav.\  {\bf 19} (2002) 5507 [arXiv:hep-th/0106100].

\bibitem{Lidsey:2003}
  J.~E.~Lidsey and N.~J.~Nunes,
  Phys.\ Rev.\  D {\bf 67}, 103510 (2003) [arXiv:astro-ph/0303168];\\
  S.~Tsujikawa, M.~Sami and R.~Maartens,
  Phys.\ Rev.\  D {\bf 70}, 063525 (2004) [arXiv:astro-ph/0406078].

\bibitem{Hwang&Noh97}
  J.~c.~Hwang and H.~Noh,
  Phys.\ Rev.\ D {\bf 54}, 1460 (1996);
C.~Cartier, J.~c.~Hwang and E.~J.~Copeland,
  Phys.\ Rev.\ D {\bf 64}, 103504 (2001) [arXiv:astro-ph/0106197].

\bibitem{Ohta07}
  Z.~K.~Guo, N.~Ohta and S.~Tsujikawa,
  Phys.\ Rev.\  D {\bf 75}, 023520 (2007) [arXiv:hep-th/0610336].


\bibitem{Leith:2007}
  B.~M.~Leith and I.~P.~Neupane,
  JCAP {\bf 0705}, 019 (2007) [arXiv:hep-th/0702002].

\bibitem{Kanno:2006}
  S.~Kanno and J.~Soda,
  Phys.\ Rev.\  D {\bf 74}, 063505 (2006) [arXiv:hep-th/0604192].


\bibitem{Koivisto:2006a}
  T.~Koivisto and D.~F.~Mota,
  Phys.\ Lett.\  B {\bf 644}, 104 (2007) [arXiv:astro-ph/0606078].

\bibitem{Koivisto:2006b}
  T.~Koivisto and D.~F.~Mota,
  Phys.\ Rev.\  D {\bf 75}, 023518 (2007) [arXiv:hep-th/0609155].

\bibitem{Nojiri:2006}
  S.~Nojiri, S.~D.~Odintsov and M.~Sami,
  Phys.\ Rev.\  D {\bf 74}, 046004 (2006) [arXiv:hep-th/0605039];
  S.~Tsujikawa and M.~Sami,
  JCAP {\bf 0701}, 006 (2007) [hep-th/0608178];
G.~Cognola, E.~Elizalde, S.~Nojiri, S.~Odintsov and S.~Zerbini,
  Phys.\ Rev.\  D {\bf 75}, 086002 (2007) [arXiv:hep-th/0611198].

\bibitem{Calcagni:2006}
  G.~Calcagni, B.~de Carlos and A.~De Felice,
  Nucl.\ Phys.\  B {\bf 752}, 404 (2006) [arXiv:hep-th/0604201].

\bibitem{EspositoFarese:2004}
  G.~Esposito-Farese,
  AIP Conf.\ Proc.\  {\bf 736}, 35 (2004)
  [arXiv:gr-qc/0409081].

\bibitem{Damour:02a}
T.~Damour, F.~Piazza and G.~Veneziano,
  Phys.\ Rev.\ D {\bf 66}, 046007 (2002) [arXiv:hep-th/0205111].

\bibitem{Amendola:2005}
  L.~Amendola, C.~Charmousis and S.~C.~Davis,
  JCAP {\bf 0612}, 020 (2006) [arXiv:hep-th/0506137].

\bibitem{SDSS}
D.~J.~Eisenstein {\it et al.} [SDSS Collaboration],
Astrophys. J. {\bf 633}, 560 (2005).

\bibitem{Amendola:2007}
  L.~Amendola, C.~Charmousis and S.~C.~Davis,
  JCAP {\bf 0710}, 004 (2007) [arXiv:0704.0175].

\bibitem{Sotiriou:2006}
  T.~P.~Sotiriou and E.~Barausse,
  Phys.\ Rev.\  D {\bf 75}, 084007 (2007) [arXiv:gr-qc/0612065].


\end{thebibliography}
\end{document}